\documentclass[12pt]{JHEP3}  

\usepackage{epsfig}
\def\journal#1&#2(#3){\unskip, \sl #1\ \bf #2 \rm(19#3) }
\def\andjournal#1&#2(#3){\sl #1~\bf #2 \rm (19#3) }

\def\frac#1#2{{#1\over#2}}

\def\ket#1{|#1\rangle}

\def\d{\partial}

\def\inbar{\,\vrule height1.5ex width.4pt depth0pt}
\def\IC{\relax\hbox{$\inbar\kern-.3em{\rm C}$}}
\def\IR{\relax{\rm I\kern-.18em R}}
\def\IP{\relax{\rm I\kern-.18em P}}

\catcode`\@=11
\def\slash#1{\mathord{\mathpalette\c@ncel{#1}}}
\def\EE{{\cal E}}

\def\t{\tau}

\def\lam{\lambda}

\def\underrel#1\over#2{\mathrel{\mathop{\kern\z@#1}\limits_{#2}}}

\catcode`\@=12

\def\ket#1{\left| #1\right\rangle}

\def \sinh{{\rm sinh}}
\def \cosh{{\rm cosh}}

\newcommand{\p}{\partial}
\newcommand{\be}{\begin{equation}}
\newcommand{\ee}{\end{equation}}

\newcommand{\QB}{Q_{\rm B}}
\newcommand{\go}{g_{\rm o}}

\def\[{[}
\def\]{]}

\def\comment#1{ }

\def\draftnote#1{\ifdraft{\baselineskip2ex
                 \vbox{\kern1em\hrule\hbox{\vrule\kern1em\vbox{\kern1ex
                 \noindent \underbar{NOTE}: #1
             \vskip1ex}\kern1em\vrule}\hrule}}\fi}
\def\internote#1{\ifinter{\baselineskip2ex
                 \vbox{\kern1em\hrule\hbox{\vrule\kern1em\vbox{\kern1ex
                 \noindent \underbar{Internal Note}: #1
             \vskip1ex}\kern1em\vrule}\hrule}}\fi}
\def\al{\alpha}
\def\bt{\beta}
                
\def\dl{\delta}                

\def\inbar{\hskip.3em\vrule height1.5ex width.4pt depth0pt}
\def\IC{\relax{\inbar\kern-.3em{\rm C}}}
\def\IN{\relax{\rm I\kern-.16em N}}
\def\IQ{\relax\hbox{$\inbar$\kern-.3em{\rm Q}}}
\def\IZ{\relax{\rm Z\kern-.8em Z}}
\def\be{\begin{equation}}
\def\ee{\end{equation}}
\def\bea{\begin{eqnarray}}
\def\eea{\end{eqnarray}}

\title{A new rolling  tachyon solution of cubic string field theory}

\author{Valentina Forini\\Dipartimento di Fisica and I.N.F.N. Gruppo
Collegato di Trento, 
Universit\`a di Trento, 38050 Povo (Trento). Italia.
\email{E-mail:forini@science.unitn.it}
\thanks{Work supported by INFN of Italy.}}

\author{Gianluca Grignani\\Dipartimento di Fisica and Sezione I.N.F.N., 
Universit\`a di Perugia, Via A. Pascoli I-06123, Perugia, Italia.
\email{E-mail:grignani$@$pg.infn.it}
\thanks{Work supported in part by INFN and MIUR of Italy.}}

\author{Giuseppe Nardelli\\Dipartimento di Fisica and I.N.F.N. Gruppo
Collegato di Trento, 
Universit\`a di Trento, 38050 Povo (Trento). Italia.
\email{E-mail:nardelli@science.unitn.it}
\thanks{Work supported in part by INFN of Italy.}}

\abstract{We present a new analytic time dependent solution of cubic
string field theory at the lowest order in the level truncation scheme.
The tachyon profile we have found is a bounce in time, a $C^{\infty}$ function which represents 
an almost exact solution, with an extremely good 
degree of accuracy, of the classical equations of motion 
of the truncated string field theory. 
Such a finite energy solution describes a tachyon which at $x^0=-\infty$
is at the maximum of the potential, at later times rolls toward the stable minimum
and then  up to the other side of the potential toward the inversion point and then
back to the unstable maximum for $x^0\to+\infty$.
The energy-momentum tensor
associated with this rolling tachyon solution can be explicitly computed. 
The energy density is constant, the pressure is an even function of time which
can change sign while the tachyon rolls toward the minimum of its potential.
A new form of tachyon matter is realized which might be relevant for cosmological applications.}


\begin{document} 

\section{Introduction}

In recent years there has been great progress, particularly due to Sen, in our 
understanding of the role of the tachyon in string theory (see~\cite{Sen:2004nf}
with references to earlier works). The basic idea is that the perturbative
open string vacuum is unstable but there exists a stable vacuum 
toward which a tachyon field naturally moves.

String theory must eventually address cosmological issues and hence it is
crucial to understand the role of time dependent solutions of the theory.
The rolling tachyon~\cite{Sen:2002nu} is an example of such a solution and
in fact it has been applied to the study of tachyon driven 
cosmology, cosmological solutions describing the decaying of unstable space filling
D-branes~\cite{Gibbons:2002md,Sen:2003mv}. In the decay, the energy
density remains constant  and the pressure approaches zero from negative values
as the tachyon rolls toward its stable minimum. This form of tachyon matter
could have astrophysical consequences and  
it then seems of utmost importance to confirm its 
existence using string field theory.

The boundary states approach to the rolling tachyon is
the one that initiated the new investigation on time dependent solutions
in string theory~\cite{Sen:2002nu}. However, the understanding of the final fate 
of the unstable D-brane and the description of the time
evolution of the boundary state are still far from being complete. 
These conformal field theory methods
provide an indirect way of constructing solutions of the classical equations
of motion without knowing the effective action. A more direct
derivation of the classical solutions can be realized by explicitly constructing the
tachyon effective action. Namely one starts from a string field theory 
in which, in principle, the coupling of the
tachyon to the infinite tower of other fields associated with massive open string states
could be taken into account. String field theory should then be a natural setting 
for the study of time dependent rolling tachyon solutions.
In the boundary string field theory (BSFT) approach to string field 
theory~\cite{Witten:1992qy} a rolling
tachyon solution has been found and can be directly associated with a given
two dimensional conformal field theory~\cite{Larsen:2002wc,Minahan:2002if,Sugimoto:2002fp}. 
The relationship between the boundary state and the boundary 
string field theory approaches is in fact very explicit.

The direct approach based on the analysis of the classical equations 
of motion of bosonic open string field theory (cubic string field theory, CSFT~\cite{Witten:1985cc})
is generally believed to be equivalent to the approach based on two 
dimensional conformal field theory. This equivalence is however less than manifest
also because it is not yet known a satisfactory rolling tachyon solution of the 
cubic string field theory equations of motion even at the classical level
and at the lowest order, the $(0,0)$, in the level truncation scheme~\cite{Moeller:2000jy,Fujita:2003ex}. 
In this paper we solve this problem providing a well behaved 
(almost exact) time dependent 
solution of the lowest order equations of motion of 
cubic string field theory.
At this order
one considers only the tachyon field and the cubic string field theory action becomes
\be
S=\frac{1}{\go^2}\int d^{26}x\left(
\frac12\,  t(x)\,(\Box+1)\, t(x)
-\frac13 \lambda_c \left(\lambda_c^{(1/3)\Box}t(x)\right)^3\right) ,
\label{taction}
\ee
where the coupling $\lam_c$ has the value 
\be
\lambda_c=3^{9/2}/2^6=2.19213 \ .
\label{lam}
\ee
Considering spatially homogeneous profiles of the
form $t(x^0)$, where $x^0$ is time, the equation of motion derived from (\ref{taction}) is
\be
(\p_0^2-1)t(x^0)+\lambda_c^{1-\p_0^2/3}
\left(\lambda_c^{-\p_0^2/3}t(x^0)\right)^2=0.
\label{eom}
\ee
We have found an almost exact analytic solution 
of this equation, 
which is given by the following well defined integral~\footnote{In
what sense this is an ``almost'' exact solution will be  explained in
section 2.}
\be
t(x^0)=\frac{9\lam_c^{-5/3}}{4\sqrt{\pi}\log\lam_c } \int_0^\infty d\tau 
\left(1-2\tau^2 \right)e^{-\tau^2} \log [\cosh x^0 +\cos(4 \tau \sqrt{\log\lam_c/3})]\ .
\label{tcsft}
\ee
Being the equation of motion time reversal invariant, the solution (\ref{tcsft})
is a symmetric bounce in $x^0$, a $C^{\infty}$ function
with the appropriate boundary conditions
to describe a rolling tachyon. 
Such a constant energy density solution, in fact, describes a tachyon which at $x^0=-\infty$
is at the maximum of the potential, at later times rolls toward the stable minimum
and then  up to the other side of the potential toward the inversion point and then
back to the unstable maximum for $x^0\to\infty$.

If the decaying D-brane is coupled to closed strings
it will act as a source for closed string modes~\cite{Chen:2002fp,Okuda:2002yd,Lambert:2003zr}. A rolling tachyon is
a time dependent source which will produce closed string radiation. All the energy
of the D-brane will eventually be radiated away into closed strings. 
In the classical rolling tachyon picture described above this would correspond 
to the introduction of a friction that would eventually stop the rolling tachyon at 
the minimum of its potential. 

The profile (\ref{tcsft}) does not present all the cumbersome features found in previous works
on rolling tachyons in cubic string field theory, 
like ever growing oscillations with time and an energy-momentum tensor 
that cannot be derived~\cite{Moeller:2002vx,Yang:2002nm,Fujita:2003ex,Erler:2004sy}.
For the tachyon profile (\ref{tcsft}) in fact the associated
energy-momentum tensor can be computed explicitly.
The energy density ${\cal E}$ is constant while
the pressure $p(x^0)$ is an even function of time. 
Pressure and energy density depend on an arbitrary constant (the constant up to which the action is defined) and they 
 can be chosen for example in such a way that the Dominant Energy Condition, ${\cal E}\ge |p(x^0)|$, holds
at any instant of time.  
In this case 
the pressure $p(x^0)$ 
starts negative when the tachyon is at the unstable maximum of the potential,
at later times becomes positive, while the tachyon reaches the minimum of the potential, and finally
it goes back to its  negative starting value at $x^0=+\infty$.
By choosing
the initial energy density to be higher, however, one might even realize the situation in
which the tachyon reaches the minimum of its potential when its pressure vanishes.
The rolling tachyon matter associated to the solution has in this case an interesting equation of state
$p(x^0)=w{\cal E}$, with $w$ that  smoothly interpolates between $-1$ and $0$, while the tachyon moves from the 
maximum of its potential to the minimum~\cite{Sen:2002in}. Passed this time, however,
the pressure becomes positive until the tachyon goes again through its minimum. 
This form of tachyon matter is thus different to the one described in~\cite{Sen:2002in,Sugimoto:2002fp,Buchel:2002tj}.

In boundary string field theory  and in most of the models used to study
tachyon driven cosmology,
the stable minimum of the potential 
is taken at infinite values 
of the tachyon field~\cite{Gerasimov:2000zp,Kutasov:2000qp,Coletti:2004ri,Sen:2002in,Sugimoto:2002fp,Buchel:2002tj}.
The tachyon thus cannot roll beyond its minimum.
One of the main objections to the rolling tachyon as a mechanism for inflation 
is that reheating and creation of matter in models where the minimum 
of the potential is at $T\to\infty$ is problematic because the tachyon
field in such theory does not oscillate~\cite{Kofman:2002rh,Gibbons:2002md}.
In cubic 
string field theory the minimum of the potential is at finite values of the
tachyon field. 
Therefore, the coupling of the free theory to a Friedman-Robertson-Walker 
metric~\cite{Gibbons:2002md}, and the consequent inclusion of a Hubble friction term,
should lead from the classical solution (\ref{tcsft})
to damped oscillations around the stable 
minimum of the potential well. 
Cubic string field theory seems then 
to open new perspectives in tachyon cosmology.

The paper is organized as follows. In Sect.2 we derive the solution (\ref{tcsft})
and discuss its analytical properties.
In Sect.3 we compute the associated energy momentum tensor,
study its time dependence and discuss the tachyon matter it describes.
In the conclusions we outlook some possible checks and applications 
of the new rolling tachyon solution of cubic string field theory.

\section{The rolling tachyon solution in cubic string field theory}

The action of cubic open string field theory reads~\cite{Witten:1985cc}
\be
S=-\frac{1}{\go^2}\int\left(
\frac12
\Phi\cdot\QB\Phi + \frac13 \Phi\cdot\left(\Phi\ast\Phi\right)\right),
\label{csft}
\ee
where $\QB$ is the BRST operator, $\ast$ is the star product
between two string fields and $\Phi$ is the open string field  containing component fields
which correspond to all the states in the string Fock space.
If we consider  only the tachyon field $t(x)$ in $\Phi$,
$\ket{\Phi}=b_0 \ket{0}t(x)$, the action (\ref{csft}) becomes (\ref{taction}).
For profiles that only depend on the time $x^0$ the equation of motion derived from (\ref{csft}) is (\ref{eom})
and we shall now look for a solution to that equation. 
Our procedure is based on the idea that
Eq.(\ref{eom}) can be generalized to become a non-linear differential equation
with an arbitrary parameter $\lam$ which substitutes the fixed value (\ref{lam})
\be
(\p_0^2-1)t(x^0)+\lambda^{1-\p_0^2/3}
\left(\lambda^{-\p_0^2/3}t(x^0)\right)^2=0 .
\label{eom1}
\ee
Then $\lam$ can be treated as an \emph{evolution} parameter. 
Fixing the initial value $\lambda=1$ one can easily find an exact solution to (\ref{eom1})   
and then one can study how this solution 
evolves to different values of $\lambda$ keeping its property of being a solution of (\ref{eom1}). 
We shall find that the equation governing the evolution in $\lam$ is extremely simple
and we shall  look for a solution of (\ref{eom1}) for generic $\lam$,
setting eventually $\lam=\lam_c$ as in (\ref{lam}).

When $\lambda=1$,  Eq.(\ref{eom}) admits a particularly simple exact solution, the following bounce
\be
t(\log \lambda=0,\,  x^0)=\frac{3}{2 \cosh^2 (x^0/2)}=6 \int_0^\infty \frac{\t \cos( \t x^0)}{\sinh (\pi \t)}\,  d\t\;\;.
\label{l1}
\ee
The boundary conditions of (\ref{l1}) are such that $\p t(0,x^0) /\p x^0 =0$ at $x^0= \pm \infty$.

Now we shall interpret the solution (\ref{l1}) as the ``initial'' condition 
of an ``evolution'' equation   with respect to the ``time'' $\log \lambda$.
To find how the solution evolves we shall have to provide a careful
treatment of infinite derivative operators of the type 
\be 
q^{\p^2}=e^{\log q \, \p^2}\equiv \sum_{n=0}^\infty \frac{(\log q)^n}{n!}\partial^{2 n}\ ,
\label{def}
\ee
which act on the function $t(x^0)$ in (\ref{eom1}) when $\lam\ne 1$. These operators
play a crucial role in string field theories and related models.
We shall thus provide a possible solution to the long standing problem of how
to treat this infinite derivative operators in string field theory.

A particularly convenient  redefinition of the tachyon field that leaves invariant the 
initial condition (\ref{l1}) is
\be 
T(\log\lam,\, x^0)= \lam^{5/3 + \p_0^2/3} t(\log\lam,\, x^0)\ .
\label{T}
\ee
With this field redefinition  Eq.(\ref{eom}) transforms into the following
\be
(\p_0^2-1)T(\log\lam, x^0)+\lambda^{-2/3}
\left(\lambda^{-2\p_0^2/3}T(\log\lam, x^0)\right)^2=0.
\label{eom2}
\ee
Since the operator $\lambda^{-2\p_0^2/3}$ is defined as a power series of $\log\lam$ through Eq.(\ref{def}), 
it is natural to look for solutions of Eq.(\ref{eom2}) of the form
\be
T(\log\lam, x^0)=\sum_{n=0}^\infty \frac{(\log\lam)^n}{n!} t_n(x^0)
\label{tser}
\ee
It  is not difficult to check that at any desired order $n$ in (\ref{tser}) 
the functions $t_n (x^0)$ can \emph{always} be written as finite sums of the form
\be 
t_n(x^0)=\sum_{k=0}^n\frac{a^{(n)}_k}{\cosh^{2 k + 2} (x^0/2)}\ ,
\label{tn}
\ee
and the differential equation for the tachyon field  becomes an algebraic equation for 
the unknown coefficients $a^{(n)}_k$.
Thus, an exact solution of (\ref{eom2}) can always be obtained as a series representation.
However, in order to obtain  solutions  preserving the correct boundary conditions, 
it is mandatory to look for solutions that, although approximate, sum the {\it whole} series
(\ref{tser}) rather than to find the exact coefficients $a^{(n)}_k$ at any fixed truncation $n$ 
of the sum (\ref{tser}). In fact, it is easy to show that any truncation of  the sum (\ref{tser})
leads to solutions with wild oscillatory behavior with increasing amplitudes, whose 
physical meaning is difficult to interpret. Only the resummation of the whole series smoothens
such oscillations.

A more convenient representation of $t_n(x^0)$ alternative to (\ref{tn})  is given by
\be
t_n(x^0)=6 \int_0^\infty \frac{\t \cos( \t x^0)}{\sinh (\pi \t)} P_n(\t) \,  d\t\ ,
\label{tn1}
\ee
$P_n(\t)$ being a polynomial of even powers of $\t$ of degree $2 n$. 
This representation is particularly useful since it provides the $t_n(x^0)$ in terms of
eigenfunction of the operator $\p_0^2$.
The field redefinition (\ref{T}) was chosen in such a way that the form of the coefficients (\ref{tn1})
becomes particularly simple. This allows an 
approximate (although very accurate) resummation of the whole series (\ref{tser}). 
With this choice, in fact, the polynomials  $P_n (\t)$ simply become
\be
P_n (\t)\simeq \t^{2 n}\ 
\label{pn}
\ee
leading to the following approximate solution of  Eq.(\ref{eom2})
\be
T(\log\lam, \, x^0)=6 \int_0^\infty \frac{\t \cos( \t x^0)}{\sinh (\pi \t)} 
e^{\log\lam \, \t^2} \,  d\t\ =6 \lam^{-\p_0^2}\int_0^\infty \frac{\t \cos( \t x^0)}
{\sinh (\pi \t)} \,  d\t, \ \ \lam<1.
\label{sol1}
\ee
Note that all the $\lam$-dependence in (\ref{sol1}) is encoded in the operator 
$\lam^{-\p_0^2}$ acting on  the solution of Eq.(\ref{eom2}) with $\lam =1$. 
In fact $T(\log\lam=0,x^0)\equiv t(\log\lam=0,x^0)$ and $\lam^{-\p^2_0}$ plays the role 
of the ``evolution'' operator (with respect to the ``time'' $\log\lam$) acting 
on the initial condition $T(\log\lam=0,x^0)$,
\be
T(\log\lam,x^0)= \lam^{-\p^2_0}\, T(\log\lam=0,x^0)\ .
\label{evolution}
\ee
Clearly, the representation (\ref{sol1}) of the solution $T(\log\lam , x^0)$ is valid only for 
$\lam \in (0,1]$. In our case the physically relevant value of $\lam$ is the one given in (\ref{lam}),
 which is greater than one.
Consequently, we need an analytical continuation of the representation (\ref{sol1})
 to positive values of $\log\lam$.

Eq.(\ref{sol1}) shows that the evolution of the tachyon field  with respect to
the parameter $\log\lam$ is simply driven by the diffusion equation with (negative) unitary coefficient.
In fact (\ref{sol1}) satisfies the diffusion equation 
\be
\frac{\p T(\log\lam , x^0)}{\p \log\lam}=- \frac{\p^2 T(\log\lam , x^0)}{\p (x^0)^2}\ 
\label{diffusion}
\ee
with respect to the ``time'' variable $\log\lam$ and the ``space'' variable $x^0$, 
with ``initial'' and ``boundary'' conditions $T(0,x^0)=3/[2 \cosh^2 (x^0/2)]$, $T(\log\lam,\pm \infty)=0$. 

Now we face the problem of the analytical continuation of the representation 
(\ref{sol1}) to positive values of $\log\lam$. Setting $\tau =-i s$ in 
Eq.(\ref{sol1}), we rewrite $T$  as
\be
T(\log \lam,x^0)
=\frac{3}{i}\lam^{-\p_0^2}\int_{-i\infty}^{+i\infty}
\frac{s e^{s x^0}}{\sin(\pi s)}\,  ds\ .
\label{sol2}
\ee
In Eq.(\ref{sol2}) the integral can be closed with semi-circles at infinity
to the right or to the left depending on the sign of $x^0$.
Let us choose for instance $x^0<0$. Then (\ref{sol2}) reads
\be
T(\log \lam,x^0<0)=-6\lam^{-\p_0^2}\sum_{n=1}^{\infty}(-1)^n n e^{n x^0}\ .
 \label{sol3}
\ee
In Eq.(\ref{sol3}) one would be tempted to replace the operator $\lam^{-\p_0^2}$ with 
its eigenvalue $\lam^{-n^2}$ inside the series, namely
\be 
-6\sum_{n=1}^{\infty}(-1)^n \lam^{-n^2}n e^{n x^0}\ ,\qquad \lam>1\ ,
\label{wrong}
\ee
thus providing very easily  the required analytical continuation to the region $\lam >1$. 
However this procedure is incorrect. This is an important point, as the solutions in cubic 
string field theory (CSFT) analyzed in the recent literature~\cite{Fujita:2003ex} 
have precisely the form (\ref{wrong}). A cavalier treatment of the
infinite derivative operator $\lam^{-\p_0^2}$, however, might lead to 
the wrong conclusion that no rolling tachyon solutions exist in CSFT.

To understand why the procedure leading to (\ref{wrong}) is incorrect, note that it
would correspond to replace the operator $\lam^{-\p_0^2}$ with $\lam^{-s^2}$ in the integrand 
of  Eq.(\ref{sol2}), and then closing the integral with a semicircle at infinity 
in the half-plane ${\rm Re}s>0$. This cannot be done when the factor $\lam^{-s^2}$ 
is inserted in the integrand. The path of integration in fact, cannot be closed by \emph{any} curve at infinity, 
for any sign of $\log\lam$: if $\lam <1$ the integral would diverge at  
$s=\pm \infty$, whereas if $\lam>1$ it would diverge at $s=\pm i \infty$ and the integral 
(\ref{sol2}) could never be computed as sum of residues. Thus,  
in spite of the fact that the series in (\ref{wrong}) has infinite convergence radius 
for $\lam>1$, it does \emph{not} provide the analytical continuation of (\ref{sol1}).

Another argument which can be given to understand why (\ref{wrong}) does not reproduce the tachyon field 
for $\lam>1$ is the following. Eq.(\ref{sol1}) is manifestly even, and then all its 
odd derivatives must vanish at the 
origin $x^0=0$~\footnote{Another possibility would be that the odd derivatives of Eq.(\ref{wrong}) 
are  discontinuous at the origin. This is indeed what happens with (\ref{wrong}). 
Clearly this is unacceptable as the resulting functions 
would  not belong to the definition domain of the operator $\lam^{-\p_0^2}$.}.
This is not true for the representation (\ref{wrong}).

A possible way to overcome these difficulties, and thus to solve
the problem of how infinite derivative operators of the type $\lam^{-\p_0^2}$
can be treated, is through a Mellin-Barnes representation 
for the operator $\lam^{-\p_0^2}$,
\be 
\lam^{-\p_0^2}= \sum_{n=0}^\infty \frac{(- \log \lam)^n}{n!}\, 
\partial_0^{2 n}=\frac{1}{2 \pi i}\int_{\gamma-i\infty}^{\gamma+i\infty}\!\!\!\! ds \, 
\Gamma (-s) (\log\lam)^s\p_0^{2 s}\ ,\ \ {\rm Re}\gamma <0 \ .
\label{def2}
\ee

\begin{figure}
\begin{center}
\includegraphics[scale=0.7]{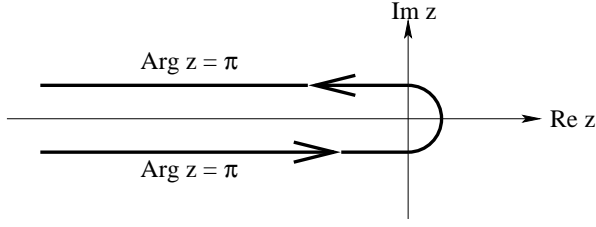}
\caption{Contour $C$}
\label{C}
\end{center}
\end{figure}

Acting with (\ref{def2}) in (\ref{sol3}), we find
\bea
T(\log \lam,x^0<0)&&=-\frac{3}{\pi i}\int_{\gamma-i \infty}^{\gamma+i\infty}
\!\!\! ds \Gamma(-s)(\log\lam)^s \sum_{n=1}^{\infty}(-1)^n n^{2s+1}e^{n x^0}\cr
&&=\frac{3 e^{x^0}}{\pi i}\int_{\gamma-i \infty}^{\gamma+i\infty}
\!\!\! ds \Gamma(-s)(\log\lam)^s \Phi(-e^{x^0},-2s-1,1)\cr
&&=\frac{12 e^{x^0}}{\sqrt{\pi}i}\int_{\gamma-i \infty}^{\gamma+i\infty}
\!\!\! ds \frac{(4\log\lam)^s}{\Gamma(-s-1/2)}\int_0^\infty\frac{d t}{t^2}
\frac{t^{-2s}}{e^{x^0}+e^t}\;\;,
\label{tlerch}
\eea
where $\Phi$ is the Lerch Transcendent defined as
\bea
\Phi( z, s,v)&&=\sum_{n=0}^\infty (v+n)^{-s} z^n\ \ , \qquad |z|<1\ ,
 \ \ \ v\ne 0, -1,-2,\dots\cr
&&=\frac{1}{\Gamma(s)}\int_0^\infty dt \ \ 
\frac{t^{s-1}e^{-(v-1)t}}{e^t-z}\ \ ,
%%\ \ \ & {\rm Re}v>0,\ \  or\ |z|\leq 1\ ,\ z\neq 1,\ {\rm Re}s>0
%%  or\ \ z=1\ ,  \{\rm Re}s>1
\label{lerch}
\eea
and the last equation in (\ref{tlerch}) follows from the integral 
representation of $\Phi$ given in (\ref{lerch}). The gamma function in
(\ref{tlerch}) can be rewritten by using the formula
\be
\frac{1}{\Gamma(-s-1/2)}=\frac{1}{2\pi i}\int_C dz\ \ e^{z}z^{s+1/2}
\label{sigaro}
\ee
where $C$ is the path drawn in Fig.\ref{C}.

Thus, the integral over $s$ in (\ref{tlerch}) can be explicitely performed,
\be
\int_{\gamma-i \infty}^{\gamma+i\infty}\!\!\!ds\left(\frac{4z\log\lam}
{t^2}\right)^s=i \pi \  \delta\left[\log t-\log\left( 2
\sqrt{z\, \log\lam}\right)\right]\;\;.
\label{delta}
\ee
In turn, integration of the $\delta$-function leads to the expression
\be
T(\log\lam,x^0<0)=\frac{3}{i \sqrt{\pi \log\lam}}
\int_C dz\ \ \frac{e^{z}}{1+e^{2\sqrt{z\, \log\lam} -x^0}} \ \ .
\label{tc}
\ee
It is easily realized that the contribution to the integral (\ref{tc}) 
given by the semicircle around the origin vanishes. 
The lower and upper branches of the path $C$ are parametrized, according to the notation of Fig.\ref{C}, as 
\bea
z&=& e^{-i \pi } t -i\epsilon\ , \qquad t\in (\infty, 0)\ ,\cr
z&=& e^{i \pi } t +i\epsilon\ , \,\,\,\qquad t\in (0, \infty)\ ,
\label{z}
\eea
respectively.
 Then, by changing variable $\tau= \sqrt{t }$, the integral (\ref{tc}) can be rewritten as 
\be
\!\!  T(\log\lam,x^0<0)=\frac{6}{\sqrt{\pi \log\lam}}\int_0^\infty \!\!\!\! d\tau
e^{-\tau^2}\frac{\tau \sin(2 \tau \sqrt{\log\lam}) }
{\cosh(x^0-\epsilon)+\cos( 2 \tau \sqrt{\log\lam})}\;, \ \ \lam>1\ .
\label{teps}
\ee
Note that the hyperbolic cosine in (\ref{teps}) is always greater than one 
($x^0\le 0$ and $\epsilon>0$), preventing any singularity in the integrand.
Analogously, if we consider the case $x^0>0$ in (\ref{sol2}), we obtain for 
$T(\log\lam,x^0>0)$ an expression similar to (\ref{teps}) with $x^0-\epsilon$ 
replaced by $x^0+\epsilon$. Therefore, in any case no singularities arise and 
a representation  for the tachyon
field valid for any value of $x^0$ can be conveniently  written as
\be
T(\log\lam,x^0)=\frac{6}{\sqrt{\pi \log\lam}}\int_0^\infty \!\!\!\! d\tau
e^{-\tau^2}\frac{\tau \sin(2 \tau \sqrt{\log\lam}) }
{e^\epsilon \, \cosh(x^0)+\cos( 2 \tau \sqrt{\log\lam})}\;, \ \ \lam>1\ .
\label{troll}
\ee

Eq.(\ref{troll}) provides the required analytical continuation of (\ref{sol1}) 
to positive values of $\log\lam$. 
Note that there is no arbitrariness in the regularization of the integral (\ref{troll}), 
as the regulator $\epsilon$ directly follows from the representation (\ref{sigaro}) 
of the gamma function. This regulator is immaterial for any point $x^0\ne 0$ but it is crucial 
to prescribe the behavior at the origin. It guarantees that $T(\log\lam,x^0)\in C^\infty$ 
in a neighbour of the origin and that all the odd derivatives of (\ref{troll}) vanish at $x^0=0$.
To understand the mechanism, we can integrate by parts Eq.(\ref{troll}) keeping $\epsilon\ne 0$. 
After integration by parts, the singularities of the denominator that would appear at 
$x^0=0$ in the $\epsilon\to 0$ limit become logarithmic (integrable) singularities. 
Then the regulator $\epsilon$ can be removed, obtaining
\be
T(\log\lam,x^0)=\frac{3}{\sqrt{\pi}\log\lam } \int_0^\infty \frac{d}{d\tau} 
\left( \tau e^{-\tau^2}\right) \log [\cosh x^0 +\cos(2 \sqrt{\log\lam}\tau)]\ .
\label{regu}
\ee
Iterating the procedure, any derivative of $T$ can be written in a manifestly regular way. 
Note that, since $\epsilon$ can be eventually removed, it works 
as a prescription to define the integral (\ref{regu}) with all its derivatives.
For example,  the formula for the even derivatives of $T$ reads
\be
\frac{d^{2 n} T(\log\lam,x^0)}{d(x^0)^{2 n}}= 
\frac{3(-1)^n}{2^{2 n}\sqrt{\pi}(\log\lam )^{n+1} } \int_0^\infty \!\!\! 
\frac{d^{2 n +1}}{d\tau^{2 n + 1}} \left( \tau e^{-\tau^2}\right) \log [\cosh x^0 +\cos(2 \sqrt{\log\lam}\tau)]\ .
\label{regu2}
\ee
The representation (\ref{regu}) is defined for 
\emph{any} real value of $\log \lam$. For $\lam >1$ it provides the analytical continuation of (\ref{sol1}),  
for $\lam <1$ it is still well defined and coincides with (\ref{sol1}).
\begin{figure}
\begin{center}
\includegraphics[scale=0.5]{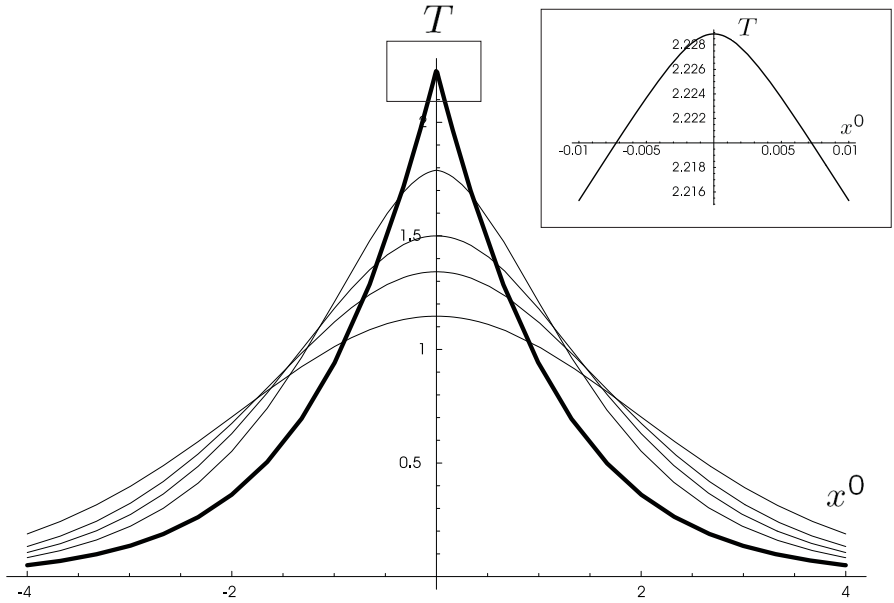}
\caption{Different profiles of the solution $T(\log\lam,x^0)$. The bold
profile refers to $\lam=\lam_c$, the remaining ones to  
$\lam_c^{1/3},1,\lam_c^{-1/3},\lam_c^{-1}$. As seen in the box, 
the behavior of the solution with $\lam=\lam_c$ is smooth at the origin.}
\label{bounce}
\end{center}
\end{figure}
The solutions (\ref{regu}) have the form of bounces, for any value of $\lambda$.
In Fig.\ref{bounce} are drawn some profiles of the solution $T$ for different values of $\lam$.
The bold profile refers to the physically relevant  value $\lambda = \lam_c$, 
the remaining ones correspond to $\lam_c^{1/3},1,\lam_c^{-1/3},
\lam_c^{-1}$. Note the manifest continuity in $\lam$
exhibited in Fig.\ref{bounce} passing form positive to negative values of $\log\lam$.

To check the level of accuracy of the approximate solution (\ref{regu}) 
we must study the action of operators of the form $q^{\p_0^2}$ on it. 
At first sight, this is a non trivial problem, as the $x^0$-dependence in (\ref{regu}) 
is not through eigenfunctions of $\p_0^2$.
Fortunately, Eq.(\ref{regu}) still satisfies the diffusion equation (\ref{diffusion}), 
as can be checked by direct inspection. Therefore, the action of the operator
$q^{\p_0^2}$ on $T(\log\lam,\, x^0)$  can be simply represented as a translation of $\log \lam$
\be 
q^{\p_0^2} T(\log\lam,\, x^0) =e^{\log q \, \p_0^2}T(\log\lam,\, x^0)=
e^{-\log q \frac{\p}{\p \log\lam}}T(\log\lam,\, x^0)=T(\log\lam - \log q,\, x^0)\ .
\label{trans}
\ee
This remarkable property can only be used thanks to the fact that we have
treated the quantity $\lam$ as a generic variable.
In particular we shall have often to make use of the following operator
\bea
\lam^{a \p_0^2} T(\log\lam,\, x^0)&=&\sum_{n=0}^\infty a^n \frac{(\log\lam)^n}{n!}
 \frac{\p^{2 n}}{\p {x^0}^{2 n}}T(\log\lam,\, x^0)\cr &=&\sum_{n=0}^\infty (-a)^n \frac{(\log\lam)^n}{n!}
 \frac{\p^n}{\p(\log\lam)^n}T(\log\lam,\, x^0)\cr
&=&T((1-a)\log\lam ,\, x^0)\ .
\label{trans2}
\eea
where in the second equality we have used the diffusion equation (\ref{diffusion}).

A quantitative estimate of the accuracy of (\ref{regu}) can be
obtained by calculating the $L_2$ norm of the left hand side
($LHS(\log\lam,\, x^0)$) of Eq.(\ref{eom2}) evaluated on the
approximate solution (\ref{regu}). If the solution of the
equation was exact the value of this norm would be zero.
Let us consider the physically
relevant case $\lam=\lam_c$.  In this case the $L_2$ norm of $LHS$
gives $||LHS||^2 = 4.636\ \cdot 10^{-8}$.

This value should be
compared with a typical scale of the problem, for instance with the
$L_2$ norm of $T$, which is $||T||^2= 2.019$. This shows the impressive
level of accuracy of the solution (\ref{regu})~\footnote{Another possibile check of the
approximation would be to write Eq.(\ref{eom2}) as $LHS=RHS$, where $LHS=(\p_0^2-1)T$,
and to consider the quantity $||LHS-RHS||^2/||LHS||^2$. The order of magnitude of this ratio
is as in (\ref{accu}).}
 \be \left(
\frac{||LHS||^2}{||T||^2}\right)_{\lam=\lam_c}\sim 2.3\ \cdot 10^{-8}\
.
\label{accu}
\ee 

 \section{Energy-momentum tensor}

 The tachyon field $t(x^0)$ appearing in the original form of the level truncated CSFT
 (\ref{taction}) is obtained by the field redefinition (\ref{T})
 applied to (\ref{regu}) with $\lam=\lam_c$. 
 Using (\ref{trans2}), one has $t(x^0)= \lam_c^{-5/3} T(\frac{4}{3}
\log\lam_c , \, x^0)$, namely 
\be
t(x^0)=\frac{9\lam_c^{-5/3}}{4\sqrt{\pi}\log\lam_c } \int_0^\infty
\frac{d}{d\tau} \left( \tau e^{-\tau^2}\right) \log [\cosh x^0 +\cos(4
\tau \sqrt{\log\lam_c/3})]\ .
\label{t}
\ee
Eq. (\ref{t}) is the analytic solution of our problem. It has the extremely good
degree of accuracy (\ref{accu})
and it does not depend on \emph{any} free parameter. In principle, 
one could try to improve the solution by
introducing some external parameter in (\ref{t}), 
but we have checked that this does not improve its accuracy.

%This is
%unusual in an approximate solution. Usually what is done in looking
%for an approximate solution is to fix some trial function depending on
%some parameters and then looking for the value of the parameters that
%better solve the equation of motion. Here is not like this. Everything
%is fixed, no free parameters are present and it uniquely depends on
%the fundamental scale in the theory, that is $\lam_c$. 

%solution (\ref{t}) is extremely stable to the introduction of free
%parameters and any slight modification of it worsen the accuracy
%(\ref{accu}).  
%This fact could lead to the intriguing possibility that
%the solution (\ref{t}) is exact (it includes also higher fields and
%interactions levels), whereas it is the equation of motion that is
%approximated. A further hint in this direction will be given below by
%studying the behavior of the energy momentum tensor.

Since we have at hand an action for the tachyon field,
the energy-momentum tensor can be calculated as usual, by first including a
metric tensor $g_{\mu \nu}$ in the action (\ref{taction}), varying the
action $S$ with respect to $g_{\mu \nu}$ and setting afterwards the
metric to be flat, $g_{\mu \nu}=\eta_{\mu\nu}$. 

It is also possible to add a constant term $-\alpha$ to the action (\ref{taction}).
This  is the only free constant we have and its choice can  be dictated by physical considerations.
In this way the tachyon potential reads
\be
V[\,t\,]=-\frac{1}{2}\, t^2+\frac{\lambda_c}{3}\,t^3 +\alpha\;\;.
\label{pot}
\ee
 
 Thus, we consider the action
\be
S=\frac{1}{\go^2}\int d^{26}x \sqrt{-g}\left(
\frac12\,  t^2\,-\frac{1}{2}\, g^{\mu \nu}\d_\mu t\,
\d_\nu t -\frac{1}{3}\lam_c \,\tilde{t}^3 -\alpha \right)\;\; ,
\label{actens}
\ee
where  $\tilde{t}=\lam_c^{\frac{1}{3}\Box}t$. The stress tensor then reads
\be
T_{\al \bt}=-\frac{2}{\sqrt{-g}}\frac{\dl S}{\dl g^{\al \bt}}\ .
\label{defin}
\ee

In varying (\ref{actens}) with respect to the metric tensor, 
one has to consider the covariant form of the D'Alembertian operator
\be
\Box = \frac{1}{\sqrt{-g}}\p_\mu\sqrt{-g}g^{\mu \nu}\p_\nu\ .
\label{box}
\ee
The variation of the operator $\lam_c^{\frac{1}{3}\Box}$ with respect
to the metric can be performed by using the following identity 
\be
\frac{\dl\lam_c^{\frac{1}{3}\Box} }{\dl g^{\alpha\beta}}=\frac{1}{3}
\log\lam_c\int_0^1 \!\! ds\, \lam_c^{\frac{1}{3}s\Box}\,
\frac{\dl\Box}{\dl g^{\alpha\beta}}\ \lam_c^{\frac{1}{3}(1-s)\Box}\ .
\label{varbox}
\ee 
An alternative way to get the variation of the infinitely many
derivatives operator $\lam_c^{\frac{1}{3}\Box}$ would be through a
power series~\cite{Moeller:2002vx,Yang:2002nm} representation of the type
(\ref{def}). However, the remarkable property (\ref{trans2}) of our solution is
particularly well suited to deal with operators of the type
$\lam_c^{\frac{1}{3}s\Box}$. In fact, their action on $T(\log\lam_c ,
x^0)$ consists in a trivial translation $\log\lam_c \to (1+\frac{1}{3}
s)\log\lam_c$.  This will permit to write the energy momentum tensor
in a simple and closed form. Most importantly, it will be written as a
bilinear in the fields $T(\log\lam_c , x^0)$ containing only
\emph{finite} derivatives. Substituting infinite derivative operators on the field $T(x^0,\log\lam_c)$ 
with the field itself, but with the parameter $\lam_c$ traslated, allows to write the
energy momentum tensor in a form analogous to that of an
ordinary (finite derivatives) field theory.
  
Taking the equation of motion (\ref{eom}) and 
Eqs.(\ref{T}),(\ref{trans2}),(\ref{actens})-(\ref{varbox}) into
account, after some integrations by parts we get the following
expression for the energy-momentum tensor
\bea
T_{\al \bt}&=&\,\lam_c^{-10/3}\left\{ \frac{}{}\dl_{\al 0}\,\dl_{\bt 0}
\left(\p_0 T({\textstyle \frac{4}{3}}\log\lam_c ,x^0)\right)^2
+g_{\al\bt}
\left[\frac{1}{2}\left(\p_0 T({\textstyle \frac{4}{3}}\log\lam_c,x^0)
\right)^2 \right.\right.\cr
&&\left.\left.
+\frac{1}{2}\left(T( {\textstyle \frac{4}{3}}\log\lam_c, x^0)\right)^2
-\frac{1}{3}\, T({\textstyle \frac{5}{3}}\log\lam_c , x^0)\,(1-\p_0^2)
 \,T(\log\lam_c , x^0)-\alpha \lam_c^{10/3}\right] \right.\cr
&&\left. -\frac{1}{3}\log\lam_c \int_0^1 ds\left[ \frac{}{}g_{\al\bt}\frac{}{}
(1-\p_0^2)\, T( {\textstyle \frac{4-s}{3}}\log\lam_c, x^0)
\,\p_0^2 T({ \textstyle \frac{4+s}{3}}\log\lam_c, x^0)\right.\right.\cr
&&\left.\left.\frac{}{}
+g_{\al\bt}( 1-\p_0^2) \,\p_0 T({\textstyle \frac{4-s}{3}}\log\lam_c , x^0)
\, \p_0 T( {\textstyle \frac{4+s}{3}}\log\lam_c, x^0)\right.\right]\cr
&&\left.\frac{}{}
+2\,\dl_{\al 0}\dl_{\bt 0}\,( 1-\p_0^2)
 \,\p_0 T({\textstyle \frac{4-s}{3}}\log\lam_c , x^0)
\, \p_0 T({ \textstyle \frac{4+s}{3}}\log\lam_c, x^0)
\frac{}{}\right\}\;\;\;.
\label{enmom}
\eea
From (\ref{enmom}) the explicit form of the energy density ${\cal E}(x^0)=T_{00}$ and  
the pressure $p(x^0)=T_{11}$ can be obtained
\bea
{\cal E}(x^0)&=&\!\lam_c^{-10/3}\left\{\frac{1}{2}\left(\p_0 T({\textstyle \frac{4}{3}}\log\lam_c ,x^0)\right)^2
\!\!-\frac{1}{2}\left(T( {\textstyle \frac{4}{3}}\log\lam_c, x^0)\right)^2\right.\cr
&& 
\left. +\frac{1}{3} T({\textstyle \frac{5}{3}}\log\lam_c , x^0)(1-\p_0^2)
 T(\log\lam_c , x^0)
+\alpha \lam_c^{10/3}\right.\cr
&& \left. 
-\frac{1}{3}\log\lam_c \int_0^1 ds \left[\frac{}{}
(1-\p_0^2)\, T( {\textstyle \frac{4-s}{3}}\log\lam_c, x^0)
\,\p_0^2 T({ \textstyle \frac{4+s}{3}}\log\lam_c, x^0)\right.\right.\cr
&&\left.\left. \frac{}{}
-( 1-\p_0^2)\, \p_0 T({\textstyle \frac{4-s}{3}}\log\lam_c , x^0)
\,\p_0 T( {\textstyle \frac{4+s}{3}}\log\lam_c, x^0)\right]\right\}\;\;,
\label{T00}
\eea

\bea
p(x^0)&=&\!\lam_c^{-10/3}\left\{\frac{1}{2}\left(\p_0 T({\textstyle \frac{4}{3}}\log\lam_c ,x^0)\right)^2
\!\!+\frac{1}{2}\left(T( {\textstyle \frac{4}{3}}\log\lam_c, x^0)\right)^2\right.\cr
&& 
\left. -\frac{1}{3} T({\textstyle \frac{5}{3}}\log\lam_c , x^0)(1-\p_0^2)
 T(\log\lam_c , x^0)
-\alpha \lam_c^{10/3}\right.\cr
&& \left. 
-\frac{1}{3}\log\lam_c \int_0^1 ds \left[\frac{}{}
(1-\p_0^2)\, T( {\textstyle \frac{4-s}{3}}\log\lam_c, x^0)
\,\p_0^2 T({ \textstyle \frac{4+s}{3}}\log\lam_c, x^0)\right.\right.\cr
&&\left.\left. \frac{}{}
+( 1-\p_0^2)\, \p_0 T({\textstyle \frac{4-s}{3}}\log\lam_c , x^0)
\,\p_0 T( {\textstyle \frac{4+s}{3}}\log\lam_c, x^0)\right]\right\}\;\;.
\label{T11}
\eea

Even if from (\ref{T00}) the energy density seems to depend strongly
on time, its plot will show that ${\cal E}(x^0)$ is actually a
constant.  The energy density is conserved and is always identical to
the chosen height of the maximum of the potential, ${\cal
E}=\alpha$. The pressure $p(x^0)$ is an even function of $x^0$, it has
the shape of a bounce in time asymptotically reaching the value
$-\alpha$.  Thus, increasing the value of $\alpha$ in (\ref{pot}), the
energy grows and the pressure lowers of the same amount.

The choice of $\alpha$ can strongly influence the physical picture
 described by the solution (\ref{t}).  However, there are some
 features that are independent on this choice, namely the qualitative
 description of the tachyon motion and the asymptotic equation of
 state, which is always $p\sim -\EE$ at $x^0\to \pm \infty$.

\begin{figure}
\begin{center}
\includegraphics[scale=0.5]{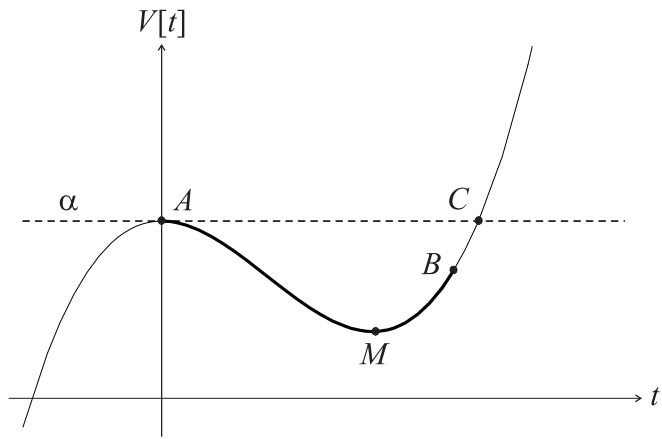}
\caption{The tachyon potential $V[t]$. 
The bold part of $V[t]$ refers to the motion $A\to M\to B\to M\to A$ of the classical solution $t(x^0)$.}
\label{potential}
\end{center}
\end{figure}

Consider the time evolution of the solution, 
 Eq.(\ref{eom}) (or (\ref{eom2}))
only admits even solutions and therefore the asymptotical states at
$x^0\to \pm \infty$ must coincide.  The motion is shown in Fig.\ref{potential}. At $x^0=-\infty$ the tachyon stays
on the maximum $A$ of the potential $V[t]$ (unstable vacuum). Since it
has no kinetic energy, its energy density - that will be conserved
during all its time evolution - is just $V[0]=\alpha$. The
pressure is negative ($p=-\alpha$), forcing the tachyon to roll towards the minimum.
As time evolves, the tachyon rolls  and at  $x_M^0=-0.144576$ reaches the 
minimum $M$ of the potential taking
 the value $t(x^0_M)=1/\lam_c$. Here the kinetic energy is maximal.
 Since ${\cal E}$ is conserved and the system is \emph{classical}, the
 tachyon cannot stop its motion and proceeds to an inversion
 point. This happens at $x^0=0$, that corresponds to $B$ in
 Fig.\ref{potential}. Note that the value of the potential at the
 inversion point $B$ is lower than the value taken in $A$, still the energy being conserved. 
 This is because the interaction felt by the
 tachyon is not described by $V[t]$, as the cubic term in the
 interaction is ``dressed'' by the kinematical factor
 $\lam_c^{-\p_0^2/3}$ (see (\ref{taction})).  This ``dressing'' is most 
 significative when the acceleration is maximal, that is precisely at
 the inversion point $B$. This is the reason why the tachyon does not
 reach the point $C$ in Fig.\ref{potential}.  For $x^0>0$ the
 tachyon inverts its motion, passing again through the minimum and
 asymptotically reaching the unstable maximum $A$ at $x^0\to +\infty$, where again $p\sim -\EE=-\alpha$.
\begin{figure}
\begin{center}
\includegraphics[scale=0.65]{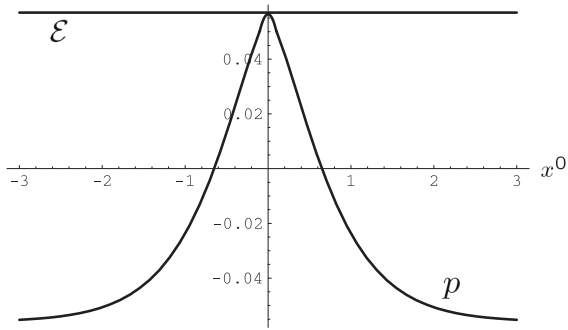}
\caption{Energy density and pressure. The value of $\alpha = 0.056$ is the minimum required to 
guarantee the DEC for any value of $x^0$. }
\label{enpres}
\end{center}
\end{figure}
As already mentioned, different choices of $\alpha =\EE$ simply raise
or lower the profile of the pressure, that maintains the shape of a
bounce.  However, different values of $\alpha$ can describe different
physical scenarios. Among all the possible choices, at least three
deserve consideration.

We can fix $\alpha$ in such a way the Dominant Energy Condition (DEC)
$\EE \ge |p(x^0)|$ holds for any value of $x^0$. This can be realized
by choosing $\alpha \ge 0.056$. In the limiting case $\alpha = 0.056$
the energy density is tangent to the pressure at the origin $x^0=0$
. At this time the equation of state $\EE \sim |p(x^0)|$ describes
\emph{stiff} matter. This is the case displayed in Fig.\ref{enpres}.

Other interesting choices can be obtained by fixing the physical
properties of the matter distribution at the minimum $M$ of the
potential. One could require that the
tachyon describes \emph{dust} when reaches $M$. Thus, by imposing
$p(x^0_M)=0$, one gets $\alpha = 0.103$.  With this choice the DEC
obvioulsy holds and the tachyon matter has the interesting equation of
state $p(x^0)=w{\cal E}$, with $w$ that smoothly interpolates between
$-1$ and $0$ while the tachyon moves from the maximum to the minimum
of its potential. Precisely as in the tachyon matter considered by Sen.
The motion however continues passed the minimum of the potential and the
pressure becomes positive. This form of tachyon matter is thus different to 
the one described in~\cite{Sen:2002in,Sugimoto:2002fp,Buchel:2002tj}.

Another intersting scenario is realized by requiring that the DEC, ${\cal
E}\ge |p(x^0)|$, holds in $x^0\in(-\infty,-x^0_M)$, {\it i.e.}
during the rolling $A\to M$ from the unstable maximum to the stable
minimum of the potential.  This is obtained by requiring that
${\cal E}= |p(x_M^0)|$, which gives $\alpha=0.051$.  Remarkably, the
choice $\alpha =0.051$ reproduces the brane tension (that in this units
is $1/(2 \pi^2)$) within the $99\%$ of accuracy. 
This might be an indication that the
solution we found might be the exact solution of the tachyon equation 
obtained by keeping into account also higher level fields.  The equation
we studied is certainly approximated, we wonder if the solution might be exact.
In fact, since $\alpha$ just gives the height of the maximum of the potential,
a natural choice for it would be the one that sets to zero the minimum
of the potential. In this case, when coupled to gravity, the potential
would not produce a cosmological constant term when the tachyon is at
the minimum.  At the $(0,0)$ level truncation we are considering, such
a constant is $1/(6 \lam_c^2)$ (which is the $68 \%$ of the brane
tension).  When all the higher level fields are taken
into account, the depth of the ``effective'' potential increases and
the constant that sets to zero the minimum of the potential should
reproduce the $D$-brane tension $1/(2 \pi^2)$.  Thus, the DEC request
naturally selects the correct depth of the potential when all
levels are included.

\section{Conclusions}

In this paper we have shown that Cubic String Field Theory (CSFT) at the lowest order in 
the level truncation scheme has
a classical rolling tachyon solution. 
This form of tachyon matter could have cosmological consequences.
Having proven its existence directly from
cubic string field theory, at least at this level of approximation, seems 
to provide a solid theoretical basis to tachyon driven cosmology.

Some interesting questions are raised by the CSFT rolling tachyon solution.
\begin{itemize}

\item
What will happen to the rolling solution if we include higher level fields and higher
powers of the tachyon field effective action? This problem should certainly be
studyed since the level (0,0) is quite a crude approximation 
that does not keep into account interactions of the tachyon with higher string modes.
At least at the classical level  this analysis is doable and interesting.
The profile we found is an extremely good approximation of the level (0,0) equations of motion,
and one wonders if the inclusion of higher level fields might just lead to an improvement 
of this approximation.  The equation
we studied is certainly approximated, we wonder if the solution might be exact.
Would the diffusion equation (\ref{diffusion}) still hold?

\item
It would be interesting to consider the coupling of the decaying D-brane,
described by the rolling solution, to
closed strings and study the emission of closed string from it.
It would in particular be interesting to see if this would cause 
damped oscillations around the minimum or it might lead to a friction
term that would just stop the rolling tachyon at the stable minimum of the potential.

\item
In order to provide a possible cosmological model it would be inconsistent not to
take into account effects of gravity during the decaying process. The coupling
of the cubic string field theory action to a Friedman-Robertson-Walker type
metric is a formidable task because of the D'Alambertians operators
in curved space that would appear in the action. If one could
still assume the validity of the diffusion equation (\ref{diffusion}),
this task could be, however, extremely simplifyed.
This might provide an alternative to the Born-Infeld type effective action that 
has been so extensively used in the study of tachyon driven 
cosmology~\cite{Gibbons:2002md,Kofman:2002rh,Frolov:2002rr,Sen:2003mv,Garousi:2004uf}.
Cubic string field theory certainly provides a tachyon effective action that correctly
describes tachyon physics~\cite{Taylor:2002fy,Taylor:2003gn} 
and it is derived from first principles.
In any case it should be at least possible to study the gravitational
effects generated by the energy-momentum tensor of the rolling tachyon solution
we have computed here, and see what kind of equations for the scale factor
this will produce.

\item
The relationship between the rolling solution 
found here and the known solution in Boundary String Field Theory (BSFT)
and vacuum string field theory~\cite{Fujita:2004ha,Bonora:2004kf}
is worth investigating~\cite{vgg}.
The former is also related to the boundary conformal field theory approach, 
so that if a link could be established between the CSFT solution and the
BSFT one, it should be possible to determine also the 
boundary state associated to the solution found here.
This should shed some more light on the relations 
between the two approaches to string field theory~\cite{Coletti:2004ri}.
It would be interesting to investigate also here the
spatial inhomogeneous decay~\cite{Larsen:2002wc,Fotopoulos:2003yt}.

\item
The solution found here does not contain free parameters,
thus it should be compared with the half-S-brane case~\cite{Larsen:2002wc,Lambert:2003zr} where 
the only parameter present can be set to 1 by a time translation.
The full S-brane case~\cite{Sen:2002nu,Gutperle:2002ai,Strominger:2002pc,Gutperle:2003xf}
contains instead a parameter
whose sign provides a prescription for which side of the
tachyon potential maximum the tachyon would roll. 
The CSFT solution we found does not present this possibility, the tachyon
always rolls to the ``right side'', i.e. to the
side where the tachyon potential is bounded below.

\end{itemize}

\acknowledgments

G.G. is grateful to Gianluca Calcagni and 
in particular to Erasmo Coletti for useful discussions.


\begin{thebibliography}{99}

\bibitem{Sen:2004nf}
A.~Sen,
``Tachyon dynamics in open string theory,''
arXiv:hep-th/0410103.
%%CITATION = HEP-TH 0410103;%%

\bibitem{Sen:2002nu}
A.~Sen,
``Rolling tachyon,''
JHEP {\bf 0204}, 048 (2002)
[arXiv:hep-th/0203211].
%%CITATION = HEP-TH 0203211;%%

\bibitem{Gibbons:2002md}
G.~W.~Gibbons,
``Cosmological evolution of the rolling tachyon,''
Phys.\ Lett.\ B {\bf 537}, 1 (2002)
[arXiv:hep-th/0204008];
%%CITATION = HEP-TH 0204008;%%%\cite{Gibbons:2003gb}
G.~W.~Gibbons,
%``Thoughts on tachyon cosmology,''
Class.\ Quant.\ Grav.\  {\bf 20}, S321 (2003)
[arXiv:hep-th/0301117].
%%CITATION = HEP-TH 0301117;%%

\bibitem{Sen:2003mv}
A.~Sen,
``Remarks on tachyon driven cosmology,''
arXiv:hep-th/0312153.
%%CITATION = HEP-TH 0312153;%%}

\bibitem{Witten:1992qy} E.~Witten,
``On background independent open string field theory,''
Phys.\ Rev.\ D {\bf 46}, 5467 (1992) [arXiv:hep-th/9208027];
E.~Witten,
``Some computations in background independent off-shell string theory,''
Phys.\ Rev.\ D {\bf 47}, 3405 (1993) [arXiv:hep-th/9210065];
S.~L.~Shatashvili,
``Comment on the background independent open string theory,''
Phys.\ Lett.\ B {\bf 311}, 83 (1993) [arXiv:hep-th/9303143];
S.~L.~Shatashvili,
``On the problems with background independence in string theory,''
[arXiv:hep-th/9311177].
%%CITATION = HEP-TH 9311177;%%.

%\cite{Larsen:2002wc}
\bibitem{Larsen:2002wc}
F.~Larsen, A.~Naqvi and S.~Terashima,
``Rolling tachyons and decaying branes,''
JHEP {\bf 0302}, 039 (2003)
[arXiv:hep-th/0212248].
%%CITATION = HEP-TH 0212248;%%

\bibitem{Minahan:2002if}
J.~A.~Minahan,
``Rolling the tachyon in super BSFT,''
JHEP {\bf 0207}, 030 (2002)
[arXiv:hep-th/0205098].
%%CITATION = HEP-TH 0205098;%%

\bibitem{Sugimoto:2002fp}
S.~Sugimoto and S.~Terashima,
``Tachyon matter in boundary string field theory,''
JHEP {\bf 0207}, 025 (2002)
[arXiv:hep-th/0205085].
%%CITATION = HEP-TH 0205085;%%

\bibitem{Witten:1985cc} E.~Witten,
``Noncommutative Geometry And String Field Theory,''
Nucl.\ Phys.\ B {\bf 268}, 253 (1986).
%%CITATION = NUPHA,B268,253;%%
%\cite{Witten:1992qy}

%\cite{Moeller:2000jy}
\bibitem{Moeller:2000jy}
N.~Moeller, A.~Sen and B.~Zwiebach,
``D-branes as tachyon lumps in string field theory,''
JHEP {\bf 0008}, 039 (2000)
[arXiv:hep-th/0005036].
%%CITATION = HEP-TH 0005036;%%

%~\cite{Fujita:2003ex}
\bibitem{Fujita:2003ex}
M.~Fujita and H.~Hata,
``Time dependent solution in cubic string field theory,''
JHEP {\bf 0305}, 043 (2003)
[arXiv:hep-th/0304163].
%%CITATION = HEP-TH 0304163;%%

\bibitem{Chen:2002fp}
B.~Chen, M.~Li and F.~L.~Lin,
``Gravitational radiation of rolling tachyon,''
JHEP {\bf 0211}, 050 (2002)
[arXiv:hep-th/0209222].

\bibitem{Okuda:2002yd}
T.~Okuda and S.~Sugimoto,
``Coupling of rolling tachyon to closed strings,''
Nucl.\ Phys.\ B {\bf 647}, 101 (2002)
[arXiv:hep-th/0208196].
%%CITATION = HEP-TH 0208196;%%

%\cite{Lambert:2003zr}
\bibitem{Lambert:2003zr}
N.~Lambert, H.~Liu and J.~Maldacena,
``Closed strings from decaying D-branes,''
arXiv:hep-th/0303139.
%%CITATION = HEP-TH 0303139;%%

%\cite{Moeller:2002vx}
\bibitem{Moeller:2002vx}
N.~Moeller and B.~Zwiebach,
``Dynamics with infinitely many time derivatives and rolling tachyons,''
JHEP {\bf 0210}, 034 (2002)
[arXiv:hep-th/0207107].
%%CITATION = HEP-TH 0207107;%%

%\cite{Yang:2002nm}
\bibitem{Yang:2002nm}
H.~t.~Yang,
``Stress tensors in p-adic string theory and truncated OSFT,''
JHEP {\bf 0211}, 007 (2002)
[arXiv:hep-th/0209197].
%%CITATION = HEP-TH 0209197;%%

\bibitem{Erler:2004sy}
T.~G.~Erler,
``Level truncation and rolling the tachyon in the lightcone basis for open
%string field theory,''
arXiv:hep-th/0409179.
%%CITATION = HEP-TH 0409179;%%}.

\bibitem{Sen:2002in}
A.~Sen,
``Tachyon matter,''
JHEP {\bf 0207}, 065 (2002)
[arXiv:hep-th/0203265];
A.~Sen,
``Field theory of tachyon matter,''
Mod.\ Phys.\ Lett.\ A {\bf 17}, 1797 (2002)
[arXiv:hep-th/0204143].
%%CITATION = HEP-TH 0204143;%%

\bibitem{Buchel:2002tj}
A.~Buchel, P.~Langfelder and J.~Walcher,
%``Does the tachyon matter?,''
Annals Phys.\  {\bf 302}, 78 (2002)
[arXiv:hep-th/0207235].
%%CITATION = HEP-TH 0207235;%%}.

%\cite{Gerasimov:2000zp}
\bibitem{Gerasimov:2000zp} A.~A.~Gerasimov and S.~L.~Shatashvili, ``On
exact tachyon potential in open string field theory,'' JHEP {\bf
0010}, 034 (2000) [arXiv:hep-th/0009103];

\bibitem{Kutasov:2000qp} D.~Kutasov, M.~Marino and G.~W.~Moore,
``Some exact results on tachyon condensation in string field theory,''
JHEP {\bf 0010}, 045 (2000) [arXiv:hep-th/0009148].
%%CITATION = HEP-TH 0009148;
%\cite{Tseytlin:2000mt}


\bibitem{Coletti:2004ri}
E.~Coletti, V.~Forini, G.~Grignani, G.~Nardelli and M.~Orselli,
``Exact potential and scattering amplitudes from the tachyon non-linear
beta-function,''
JHEP {\bf 0403}, 030 (2004)
[arXiv:hep-th/0402167].
%%CITATION = HEP-TH 0402167;%%


\bibitem{Kofman:2002rh}
L.~Kofman and A.~Linde,
%``Problems with tachyon inflation,''
JHEP {\bf 0207}, 004 (2002)
[arXiv:hep-th/0205121].
%%CITATION = HEP-TH 0205121;%%
%\cite{Chen:2002fp}

%%CITATION = HEP-TH 0209222;%%

\bibitem{Frolov:2002rr}
A.~V.~Frolov, L.~Kofman and A.~A.~Starobinsky,
%``Prospects and problems of tachyon matter cosmology,''
Phys.\ Lett.\ B {\bf 545}, 8 (2002)
[arXiv:hep-th/0204187].
%%CITATION = HEP-TH 0204187;%%

%\cite{Garousi:2004uf}
\bibitem{Garousi:2004uf}
M.~R.~Garousi, M.~Sami and S.~Tsujikawa,
%``Inflation and dark energy arising from rolling massive scalar field on the
%D-brane,''
Phys.\ Rev.\ D {\bf 70}, 043536 (2004)
[arXiv:hep-th/0402075].
%%CITATION = HEP-TH 0402075;%%

\bibitem{Taylor:2002fy}
W.~Taylor,
``A perturbative analysis of tachyon condensation,''
JHEP {\bf 0303}, 029 (2003)
[arXiv:hep-th/0208149].
%%CITATION = HEP-TH 0208149;%%\cite{taylor}

\bibitem{Taylor:2003gn}
W.~Taylor and B.~Zwiebach,
%``D-branes, tachyons, and string field theory,''
arXiv:hep-th/0311017.
%%CITATION = HEP-TH 0311017;%%. 
%\cite{Fujita:2004ha}
\bibitem{Fujita:2004ha}
M.~Fujita and H.~Hata,
%``Rolling tachyon solution in vacuum string field theory,''
Phys.\ Rev.\ D {\bf 70}, 086010 (2004)
[arXiv:hep-th/0403031].
%%CITATION = HEP-TH 0403031;%%

\bibitem{Bonora:2004kf}
L.~Bonora, C.~Maccaferri, R.~J.~Scherer Santos and D.~D.~Tolla,
%``Exact time-localized solutions in vacuum string field theory,''
arXiv:hep-th/0409063.
%%CITATION = HEP-TH 0409063;%%

\bibitem{vgg}V. Forini, G.~Grignani and G.~Nardelli, in preparation. 

\bibitem{Fotopoulos:2003yt}
A.~Fotopoulos and A.~A.~Tseytlin,
``On open superstring partition function in inhomogeneous rolling tachyon
%background,''
JHEP {\bf 0312}, 025 (2003)
[arXiv:hep-th/0310253].
%%CITATION = HEP-TH 0310253;%%Niarchos:2004rw}


\bibitem{Gutperle:2002ai}
M.~Gutperle and A.~Strominger,
``Spacelike branes,''
JHEP {\bf 0204}, 018 (2002)
[arXiv:hep-th/0202210].
%%C%\cite{Strominger:2002pc}

\bibitem{Strominger:2002pc}
A.~Strominger,
``Open string creation by S-branes,''
arXiv:hep-th/0209090.
%%CITATION = HEP-TH 0209090;%%

%\cite{Gutperle:2003xf}
\bibitem{Gutperle:2003xf}
M.~Gutperle and A.~Strominger,
``Timelike boundary Liouville theory,''
Phys.\ Rev.\ D {\bf 67}, 126002 (2003)
[arXiv:hep-th/0301038].
%%CITATION = HEP-TH 0301038;%%




\end{thebibliography}
\end{document}